\RequirePackage[T1]{fontenc}

\documentclass[conference,a4paper]{IEEEtran}
\IEEEoverridecommandlockouts
\usepackage[english]{babel}
\usepackage{cite}
\usepackage[hidelinks]{hyperref} 
\usepackage{amsmath,amssymb,amsfonts}
\usepackage{siunitx}
\usepackage{algorithmic}
\usepackage{graphicx}
\usepackage{textcomp}
\usepackage{xcolor}
\usepackage{siunitx}
\usepackage{booktabs}
\usepackage{comment}

\def\BibTeX{{\rm B\kern-.05em{\sc i\kern-.025em b}\kern-.08em
    T\kern-.1667em\lower.7ex\hbox{E}\kern-.125emX}}
\begin{document}

\title{ Predicting BESS Degradation with Uncertainty Quantification: A Probabilistic Framework for Battery Energy Storage Systems 
\thanks{Corresponding author's e-mail: melina.graner@tum.de}
\thanks{This research is funded by the German Federal Ministry for Economic Affairs and Energy (BMWE)  via the research project BattLifeBoost (grant number 03EI4068C).}
\thanks{This work has been submitted to the IEEE for possible publication. Copyright may be transferred without notice, after which this version may no longer be accessible.}
    }

\author{
    Melina Graner\textsuperscript{1,2},      
    Holger Hesse\textsuperscript{2},
    Andreas Jossen\textsuperscript{1} \\[0.1cm]
    \textsuperscript{1}Technical University of Munich,
    TUM School of Engineering and Design, \\
    Department of Energy and Process Engineering,
    Chair of Electrical Energy Storage Technology, Germany\\
    \textsuperscript{2}Kempten University of Applied Sciences; Institute for Energy and Propulsion Technologies, Kempten, Germany \\
}
\maketitle
\begin{abstract}
Accurate and uncertainty-aware prediction of battery degradation is essential for the reliable operation and lifecycle management of energy storage systems, yet traditional
deterministic models fail to capture the inherent uncertainty in degradation processes. This study introduces a framework for probabilistic battery state-of-health prediction.
The framework leverages deep learning models to generate predictive distributions for capacity loss, conditioned on stress factors. Uncertainty is propagated through stochastic degradation trajectories, enabling robust predictions even under dynamic operating conditions. A key advancement is the framework's scalability to full-system data: by integrating cell-level predictions with system topology and real-world operational variability, it provides probabilistic estimates for entire battery energy storage systems. The approach is tested using multi-year field data from residential storage systems, demonstrating its ability to mimic system-level degradation behavior. The framework predicts SOH degradation with 95\% prediction intervals that align well with remaining capacity measurements performed on the field system. This work bridges the gap between laboratory test derived battery cell aging models and full-system operational data evaluation for degradation estimation, offering a practical tool for data-driven asset management in modern energy systems.
\end{abstract}


\section{Introduction}
Battery Energy Storage Systems (BESS) are a cornerstone of the energy transition, providing essential flexibility for integrating variable renewable energy sources and ensuring grid stability \cite{Hannan.2021}. Accurate battery modeling is critical, as operation reliability and revenues depend on precise quantification of dispatchable energy content \cite{Reniers.2021}. 
Lithium-ion batteries (LIB) have emerged as the dominant technology for applications ranging from consumer electronics to electric vehicles (EVs) and stationary storage systems, thanks to their high energy and power densities. However, cyclization induces irreversible chemical reactions and structural changes, gradually reducing capacity and power capabilities.\cite{Birkl.2017} Battery systems typically consist of multiple cells grouped into subunits or modules, connected in series and parallel configurations. These subunits age at different rates due to factors such as manufacturing variability, unbalanced cell usage and thermal imbalances, leading to heterogeneous operational behavior~\cite{Barbers.2024}. This poses significant challenges for risk-based planning and operation of power systems, where uncertainty in battery performance can impact grid stability, economic dispatch, and reliability. Probabilistic approaches are essential to quantify and manage these risks, enabling operators to make informed decisions in dynamic and uncertain environments. 

Historically, battery degradation was modeled using empirical or semi-empirical formulations which link capacity loss to operating conditions and known aging mechanisms. Fundamental works by Naumann et al.~\cite{Naumann.2018, Naumann.2020} established relationships between degradation and stress factors such as temperature, charge/discharge rates, and depth-of-charge (DOC). Furthermore pseudo-two-dimensional (P2D) electrochemical models exist that describe the internal electrochemical mechanisms of batteries~\cite{Chen.2024}. They come with high computational costs and require careful parametrization. \cite{Rosewater.2019,LozanoRuiz.2025, Hu.2020} Meanwhile data-driven techniques, particularly machine learning (ML), have gained prominence for battery diagnostics and prognostics as they enable efficient application, and their retraining or recalibration requires less manual parametrization effort. Combining physical insight with ML can further improve predictive performance, especially under variable operating conditions. However, most ML-based approaches produce deterministic, pointwise predictions of battery capacity, failing to capture the uncertainty inherent in real-world operation.~\cite{Thelen.2024} This uncertainty can arise and accumulate from multiple sources including dynamic stress states during operation, cell-to-cell variability (e.g., manufacturing differences) and environmental conditions. Ignoring it limits the reliability of maintenance planning, safety assessment, and lifecycle management decisions.

To address this gap, this study introduces an open-source, probabilistic framework for battery aging prediction. Using deep learning models trained on $\mathrm{LiFePO_4}$ (LFP) battery aging datasets, the framework generates uncertainty-aware predictions of cumulative capacity loss and state-of-health (SOH), including prediction intervals that account for uncertainty. By generating particle-based degradation trajectories, the framework gives operators greater insight into the uncertainty surrounding the timing of critical SOH-threshold crossings, manage resources to avoid premature replacements or unexpected failures, and make early, informed decisions regarding retrofitting, repurposing, or potential second-life applications. This reduces operational costs, extends asset lifetimes, and enhances sustainability by minimizing premature replacements. While the focus is on stationary storage systems, the principles are broadly applicable to all battery-powered devices, including electric vehicles (EVs) and consumer electronics, where reliable lifetime estimation and safety are critical. This work provides a scientifically robust and practically applicable tool for sustainable, lifecycle-aware battery management. The key contributions of this paper are:
\begin{itemize}
    \item A probabilistic regression framework incorporating an approach for physically consistent capacity loss propagation across dynamic operation conditions.
    \item Application on real-world field data, expanding previous cell-level degradation predictions towards (multi-cell) system level applicability. 
    \item Uncertainty quantification that enables the estimation of SOH-threshold crossing likelihood for field operation of BESS.    
\end{itemize}

\section{Simulation Framework}
\label{sec:framework}
This study presents an open-source framework for probabilistic battery degradation modeling implemented in the Git repository "BattProDeep" \cite{BattProDeep.github}. At the core of the framework are two probabilistic regression models that predict total capacity loss as a conditional Gaussian distribution:
\begin{equation}
L \mid v \sim \mathcal{N}\bigl(f_\mu(v), f_\sigma^2(v)\bigr).
\end{equation}
Both the mean $f_\mu(v)$ and variance $ f_\sigma^2(v)$ are learned from the  training data ($v$ is the vector of stress features). The training dataset originates from an aging study of commercial $\mathrm{LiFePO_4}$/ graphite cells (Sony/Murata), as detailed by Naumann et al. \cite{Naumann.2018, Naumann.2020}. This dataset is exceptionally rich for model training, as it includes 17 test points for calendar aging and 19 test points for cyclic aging, covering a wide range of stress factors. Notably, three cells were aged for 885 days under identical conditions for each test point, providing a statistical basis for capturing aleatoric uncertainty in degradation behavior. Epistemic uncertainty regarding the mean $f_\mu(v)$ model output is obtained via an ensemble of bootstrap-trained models across the different test points and time. The models were validated for accuracy using dynamic load profiles of stationary battery storage systems. For further information on feature extraction, model architecture, training procedures and the validation, refer to \cite{Heidarabadi.2024} and the open-source repository \cite{BattProDeep.github}.
\begin{figure}[tb]
    \centering
    \includegraphics[width=1\columnwidth]{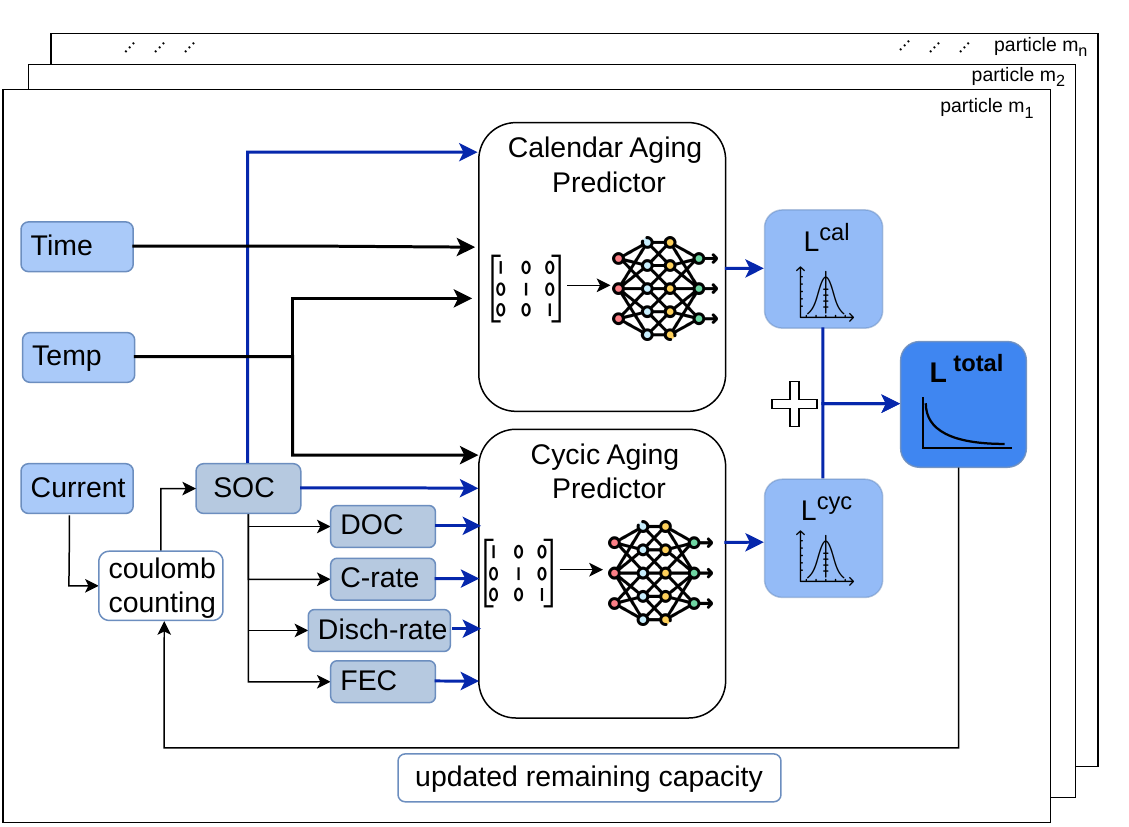} 
    \caption{The proposed framework that applies the probabilistic ML-models for the prediction of battery aging trajectories under varying stress conditions. The looped approach secures the consideration of the aging state itself during the stress factor extraction. Two separate ML-models predict cumulative calendar and cyclic losses $L^{\mathrm{cal}}$ and $L^{\mathrm{cyc}}$. Total cumulative losses $L^{\mathrm{total}}$ are determined by their sum. A set of M different particle trajectories are predicted in this manner, reflecting the impact of prediction uncertainties over time.}
    \label{fig:framework}
\end{figure}
The simulation framework, illustrated in Fig.~\ref{fig:framework}, processes multi-year time-series data by dividing it into 90-day blocks. For each block, the preprocessing step extracts relevant features required for prediction. The dependency of the SOC on the remaining capacity motivates for iterative batch processing. The predicted capacity loss is used to update the remaining capacity in the battery model, enabling the computation of SOC profiles and cyclic features for subsequent blocks.
The concept of $virtual~time$ from Naumann et al.~\cite{Naumann.2018} is employed to translate capacity loss across varying stress states into a unified equivalent loss history. It represents the time that would have needed to pass to reach the total past calendar capacity loss under the present stress factors. Separate models are utilized for calendar and cyclic aging.

\subsection{Aging Predictors}
\subsubsection{Dynamic Calendar Aging Predictor}
The framework processes dynamic SOC and temperature profiles in 90-day blocks. For each predefined time window $w$ (by default one hour), feature vectors are generated, containing averaged SOC and temperature and $virtual~time$ for two successive prediction windows $i-1$ and $i$. The current $virtual~time$ ($t^*_{i}$) is calculated as: 
\begin{equation}
t^*_{i}=t^*_{i-1}+\Delta t
\end{equation}
where ($\Delta t$) is the duration of the window.

The mean total capacity loss $\mu^{L^{\mathrm{cal}}}$ and its standard deviation $\sigma^{L^{\mathrm{cal, al}}}$ are predicted for each time window $w$ using the probabilistic calendar aging model $f$: 
\begin{align}
\mu_{t_{i}}^{L^{\mathrm{cal}}} &= f_\mu \bigl(t^*_i, \overline{T}, \overline{\mathrm{SOC}} \bigr) \label{eq:16} \\
\sigma_{t_{i}}^{L^{\mathrm{cal, al}}} &= f_\sigma \bigl(t^*_i, \overline{T}, \overline{\mathrm{SOC}} \bigr) \label{eq:18} 
\end{align}
The model output uncertainty is estimated from the dispersion of the mean predictions produced by the ensemble of bootstrap-trained calendar models. Thus, the bootstrap ensemble induces an empirical predictive distribution over $\mu_{t_i}^{L^{\mathrm{cal}}}$, whose standard deviation is interpreted as epistemic model uncertainty $\sigma_{t_{i}}^{L^{\mathrm{cal, ep}}}$.
The $virtual~time$ is adjusted for the next prediction loop based on the cumulative capacity loss $L^{\mathrm{cal}}$ using an inverse virtual~time~model~$f^{-1}$.
\begin{equation}
t^*_{i-1} = f^{-1} \bigl(L^{\mathrm{cal}}, \overline{T}, \overline{\mathrm{SOC}} \bigr) \label{eq:21}
\end{equation}
This adjustment ensures a consistent transition of the degradation process across varying aging conditions.
\subsubsection{Dynamic Cyclic Aging Predictor}

The preprocessing identifies the finalization of significant charging and discharging events, known as half-cycles, and annotates features such as depth of cycle, charge/discharge rates, and cumulative Full Equivalent Cycles (FEC). For each FEC window ($w_{fec}$, by default one FEC), the cyclic stress features of all micro-cycles within that window are aggregated into a single feature vector. Feature vectors are formed for two successive prediction states, $FEC_{i-1}$ and $FEC_i$, and include the average SOC, average temperature, maximum DOC, average charge rate, average discharge rate, and cumulative FEC. The cumulative FEC is updated as:
\begin{equation}
FEC_{i} = FEC_{i-1} + \Delta FEC \label{eq:23}
\end{equation}
where $\Delta FEC$ represents incremental number of FECs identified during preprocessing. 
The mean total capacity loss $\mu^{L^{\mathrm{cyc}}}$ and its standard deviation $\sigma^{L^{\mathrm{cyc, al}}}$ are then predicted for each FEC window using the separate probabilistic cyclic aging model $g$ which accounts for additional stress factors specific to cycling:
\begin{align}
\mu_{FEC_{i}}^{L^{\mathrm{cyc}}} &= g_\mu \big(FEC_{i}, \overline{\text{C-R.}}, \overline{\text{Disc-R.}}, \overline{T}, \overline{\mathrm{SOC}}, DOC \big) \label{eq:30} \\
\sigma_{FEC_{i}}^{L^{\mathrm{cyc, al}}} &= g_\sigma \big(FEC_{i}, \overline{\text{C-R.}}, \overline{\text{Disc-R.}}, \overline{T}, \overline{\mathrm{SOC}}, DOC \big) \label{eq:32}
\end{align}
The model output uncertainty $\sigma_{FEC_{i}}^{L^{\mathrm{cyc, ep}}}$ is obtained from the ensemble of bootstrap-trained cyclic models.

\subsection{Uncertainty Quantification and Stochastic Trajectory Interpretation}
\label{sec:uncertainty}
The training dataset contains multiple degradation observations for identical input feature vectors, corresponding to cells aged under nominally identical operating conditions but exhibiting different degradation behavior. This variability can be interpreted as aleatoric uncertainty, which is irreducible and persists even for a perfectly specified model (expressed in $\sigma^{L, al}_{i}$). At the same time, uncertainty also arises from epistemic sources, including finite training data, model structure, and extrapolation beyond the observed operating regime (expressed in $\sigma^{L, ep}_{i}$).
Dynamic usage profiles accumulate the incremental losses of single windows $\Delta\mu^{L}_{i} = \mu^{L}_{i} - \mu^{L}_{i-1}$ over time to obtain a final degradation trajectory. The temporal propagation of the uncertainties on the other hand requires careful handling. A naive approach would assume statistically independent prediction-increments at each window and accumulate uncertainty by summing incremental variances over time. However, such an assumption is not appropriate for battery aging prediction. Although the state variables vary across the windows, battery degradation itself is inherently time-dependent and these state variables encode the cumulative degradation history. Degradation at a given timestep is therefore governed by the accumulated exposure of the cell to stress conditions and their potentially irreversible impacts on the material level. Furthermore, the training datasets show that variability occurs primarily between cells rather than within individual degradation trajectories. Therefore, uncertainty is modeled at the trajectory level instead of as independent timestep-wise fluctuations.

To address this, uncertainty is modeled using stochastic degradation trajectories. A set of $M$ particles is generated, each representing a plausible degradation path. In accordance with literature, for each particle $m$ two stochastic deviation parameters $\varepsilon_m, \zeta_m \sim \mathcal{N}(0,1)$ are sampled once from a standard normal distribution and held constant throughout the simulation. Separate realizations of $\varepsilon_m$ and  $\zeta_m$ are used for the calendar and cyclic aging contributions.~\cite{Paul.2013} The stochastic incremental loss for each particle $m$ is then derived as:
\begin{align}
L^{(m)}_{i-1}
&=
\mu^L_{i-1}
+
\zeta_m \cdot \sigma^{L, ep}_{i-1}
+
\varepsilon_m \cdot \sigma^{L, al}_{i-1}, \\
L^{(m)}_{i}
&=
\mu^L_{i}
+
\zeta_m \cdot \sigma^{L, ep}_{i}
+
\varepsilon_m \cdot \sigma^{L, al}_{i}, \\
\Delta L_i^{(m)}
&=
L^{(m)}_{i}
-
L^{(m)}_{i-1}.
\end{align}
To ensure that the predictions are meaningful, any negative outputs are clipped to zero, as negative capacity losses are physically not possible in the context of battery capacity loss. This construction ensures that uncertainty evolves consistently with the predicted cumulative loss, which is obtained by summation of incremental losses
\begin{equation}
L_{\mathrm{cum},i}^{(m)}
= \sum_{j=1}^{i} \Delta L_j^{(m)}
\end{equation}
where $j$ indexes successive prediction windows and from which the SOH trajectories are derived as
\begin{equation}
\mathrm{SOH}_i^{(m)}
= 1 - L_{\mathrm{cum},i}^{(m)}
\end{equation}

Prediction intervals for SOH are computed empirically from the ensemble of particle trajectories using quantiles. Instead of propagating uncertainty by repeatedly sampling independent noise or accumulating variances over time, the proposed approach represents uncertainty of battery aging prediction through persistent stochastic deviations at the trajectory level,  reflecting structural model uncertainty and cell-to-cell variability rather than temporally uncorrelated process noise. This enables physically consistent uncertainty quantification for battery aging in dynamic usage scenarios.


\section{Simulation Case Study}
\begin{figure}[htb]
    \centering
    \includegraphics[width=1\columnwidth]{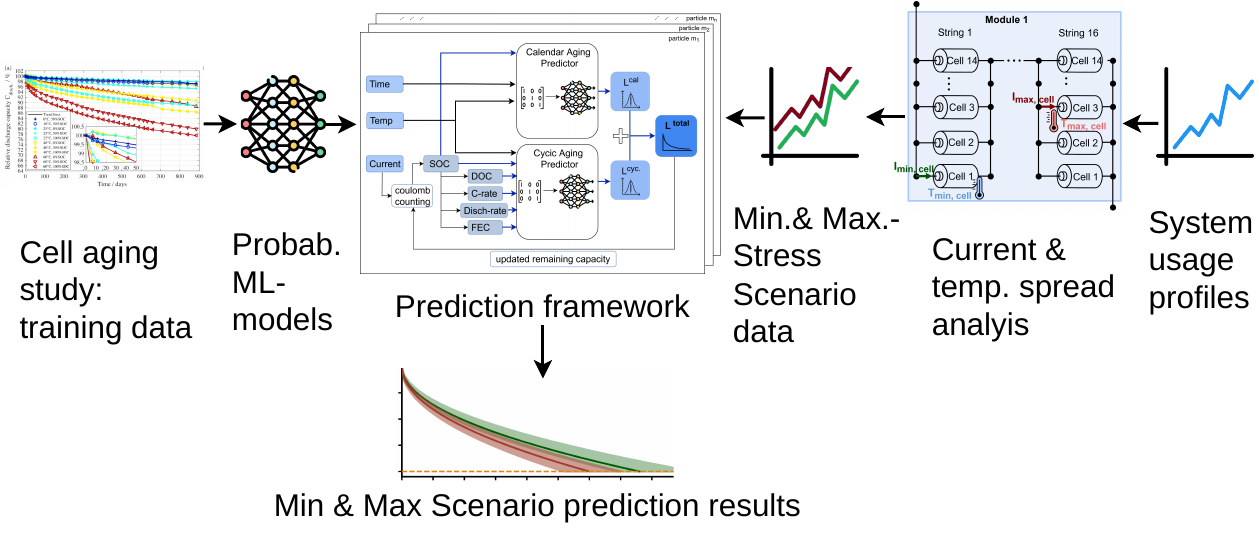} 
    \caption{Graphical overview of implementing the framework presented in section \ref{sec:framework}. The LFP cell aging datasets of Naumann et al.~\cite{Naumann.2018, Naumann.2020} are the training data for the probabilistic ML-models. The system aging behavior is estimated by combining two extreme stress scenarios that single cells in a system can experience (Minimum- and Maximum-Stress-Scenario). The prediction framework depicted in Fig.~\ref{fig:framework} processes the scenario's usage data to predict the set of aging trajectories for the Minimum- and Maximum-Stress-Scenario.}
    \label{fig:methodology}
\end{figure}

This section demonstrates the application of the probabilistic battery degradation framework introduced in Section~\ref{sec:framework} to real-world BESS field data, in order to approximate system-level degradation behavior (see  Fig.~\ref{fig:methodology}). 

\subsection{Data}
To evaluate the framework's applicability to real-world cycling data, we utilize a public dataset published by Figgener et al.~\cite{Figgener.2024}. This dataset comprises multi-year field measurements of private home storage systems in Germany. The systems, combined with residential photovoltaic  installations, were monitored capturing the system's current, voltage, power, pack housing temperature, and room temperature.
For this study, we focus our case study on a single specific system (system 14), as it employs the same $\mathrm{LiFePO_4}$/ graphite cells used to train our ML-models and further field datasets (containing temperature measurements on cell-level and current measurements on module-level) as well as field capacity test measurements performed by Figgener et al.~\cite{Figgener.2024} are available for this system, enabling benchmarking the prediction results. Since the framework presented in section \ref{sec:framework} predicts capacity loss at the cell-level, we scale the available data down to the cell-level, accounting for the BESS topology. 
\begin{figure}[htb]
    \centering
    \includegraphics[width=1\columnwidth]{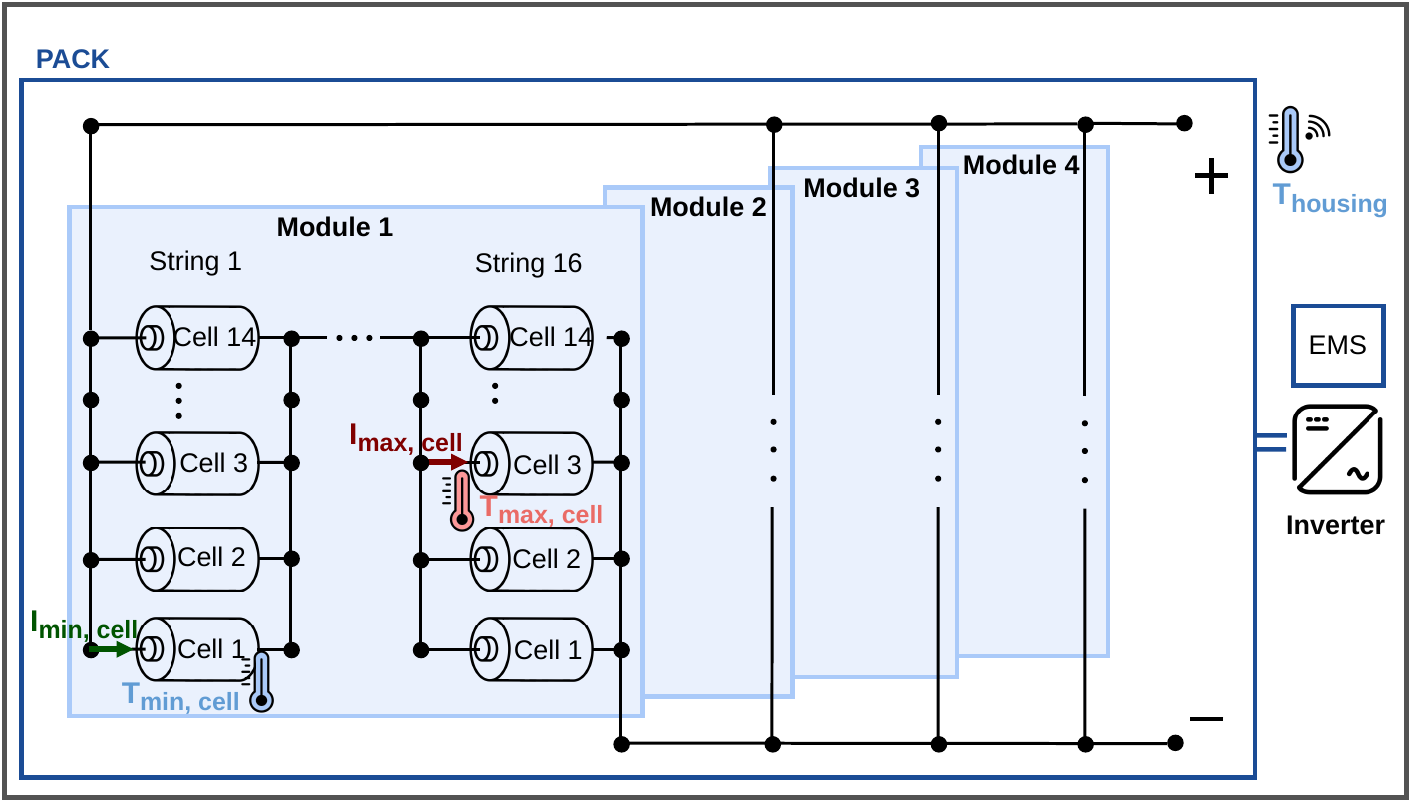} 
    \caption{Topology of the BESS: Each module is configured as 16s14p. The system contains four such identical modules. Non-uniform current and temperature distribution causes varying stress levels. Here, Cell 3 in String 16 experiences highest stress, while Cell 1 in String 1 faces the lowest stress. Different stress states lead to different aging rates of individual cells.}
    \label{fig:bess_topology}
\end{figure}

\subsection{Scaling from System- to Cell-Level}
\label{sec:celltosystem}
The system consists of four modules, each comprising a 16s14p cell configuration, totaling 896 cells with a nominal capacity of 3 Ah each. Figure~\ref{fig:bess_topology} illustrates the system topology.
Cell-level currents were estimated assuming homogeneous current sharing among the $n^{parallel}$ cells within each module. Accordingly, the measured module-level currents were divided by the number of parallel cells:
\begin{align}
I_i^{max,cell} =I_i^{max, module} / n^{parallel}\label{eq:cell_I_max}\\
I_i^{min,cell} =I_i^{min, module} / n^{parallel} \label{eq:cell_I_min}
\end{align}
Literature indicates that current imbalance in parallel-connected cells is primarily caused by non-uniform impedance, temperature, contact resistance, SOC or aging. Large current deviations reported in literature are often observed in studies designed to accentuate such effects, and are therefore not considered representative of the present commercial modules. \cite{Brand.2016,Roehrer.2025,Jocher.2024,Diao.2019, Hofmann.2018} As both the impedance spread reported for the investigated $\mathrm{LiFePO_4}$/ graphite cell\cite{Rumpf.2017} and the observed temperature gradients between cells are small (see below), equal current sharing was adopted as a first-order approximation.
The dataset of Figgener et al. provides pack housing temperature. The battery temperature sensors were attached to the outside of the battery pack or inside the BESS housing and therefore do not reflect the actual battery temperature and only serve as a rough indication. \cite{Figgener.2024} To further capture thermal variability within the pack, we use the extreme temperature values $T_i^{max,cell}$ and $T_i^{min,cell}$ in our own datasets, recorded by temperature sensors at cell surface. They show a difference between maximum and minimum cell temperatures of $1.56\,^\circ\mathrm{C}$ on average.
To account for uncertainties in scaling operational data from pack to cell-level, we considered two extreme scenarios that provide a physically motivated approximation for the system-level degradation:
\begin{itemize}
    \item Maximum-Stress Scenario:  Combines the maximum cell-level current ($I_i^{max, cell}$) with the maximum cell-level temperature ($T_i^{max, cell}$)
    \item Minimum-Stress Scenario: Combines the minimum cell-level current ($I_i^{min, cell}$) with the minimum cell-level temperature ($T_i^{min, cell}$)
\end{itemize}
Battery pack aging is expected to fall within the corridor defined by these scenarios. 

\subsection{Prediction Results and Evaluation}
\begin{figure*}[bt]
  \centering
  \includegraphics[width=\textwidth]{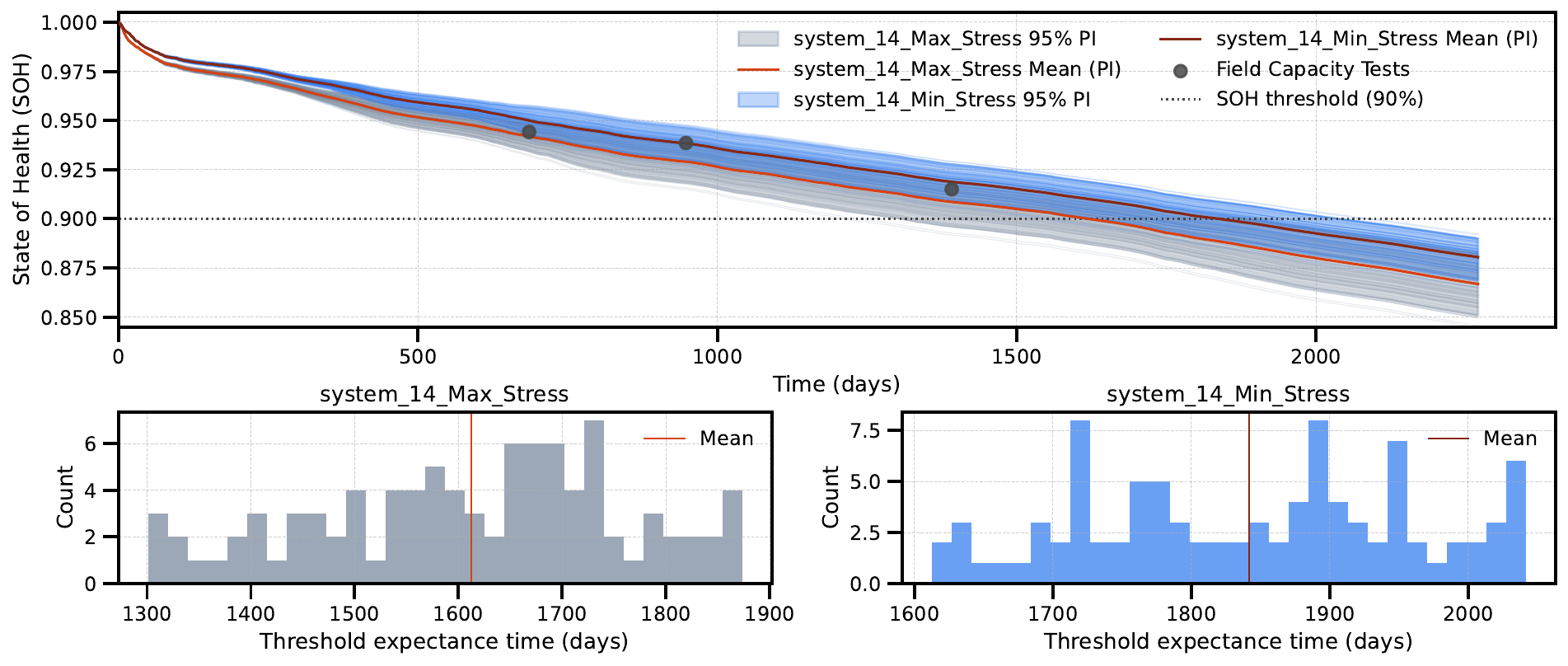}
  \caption{Top: SOH trajectories for the Maximum-Stress and Minimum-Stress Scenario that describe the system-level degradation. The 95\% prediction intervals (shaded area) captures the divergence of particle trajectories over time. All three field capacity measurements converted to SOH values (dark dots) fall within the prediction intervals. Bottom: Distribution of predicted SOH-threshold crossing days under Maximum-Stress and Minimum-Stress Scenarios. We assumed an SOH threshold of 90\% SOH here.}  
  \label{fig:Figgener14results}
\end{figure*}
In this section, we present the prediction results under both stress scenarios. For both the Maximum-Stress and Minimum-Stress Scenario, we sample $M$ = 99 particle trajectories. Each particle trajectory represents a plausible degradation path. They diverge over time due to variations in SOC profiles and input feature vectors during preprocessing. Fig.~\ref{fig:Figgener14results} displays the mean SOH trajectory for each scenario, along with the 95\% prediction interval (PI), computed empirically from the distribution of all M stochastic degradation trajectories using the 2.5th and 97.5th percentiles:
\[
\text{PI}_i = \big[ \text{SOH}_i^{(2.5\%)}, \ \text{SOH}_i^{(97.5\%)} \big]
\]
At the end of the usage profile, the Maximum-Stress Scenario predicts a final SOH of 86.68\%, while the Minimum-Stress Scenario predicts a final SOH of 88.05\%. The 95\% PIs of both scenarios evolve  differently due to variations in model output uncertainty under the different stress conditions. 

In the second, third and fourth year of operation field capacity tests were performed by Figgener et al., using the test procedure described in the supplementary information \cite{Figgener.2024}. Transformed to SOH values, they are marked as dark dots in Fig.~\ref{fig:Figgener14results}. All three of those values lie within the PIs of our predictions and close to the two predicted mean SOH trajectories. 

Probabilistic SOH predictions enable estimation of the expected degradation timeline of the battery system. To analyze the likelihood of reaching a critical SOH level, we define a 90\% SOH marker (as it falls within the range of both scenarios). The distribution of stochastic trajectories crossing this marker reflects the variability in the predicted timeline for reaching this state. Even minor differences in SOH across trajectories translate into significant variation in the anticipated intercept with the 90\% threshold line.  Fig.~\ref{fig:Figgener14results} shows the histogram of SOH-threshold crossing days across all trajectories. For the Maximum-Stress Scenario the system is predicted to reach the 90\% SOH intercept earliest after 1305 days and latest after 1870 days (1612 days for the mean trajectory). For the Minimum-Stress Scenario, the intercept is expected earliest after 1615 days and latest after 2035 days (1842 days for the mean trajectory). 
The predictions match the field capacity measurements very well, and the predicted lifetime behavior is supported within the validated operating regime (885 days and up to 2400 FECs down to 86.4\% SOH in the PV-HESS validation profile (photovoltaic-coupled home energy storage system)~\cite{Heidarabadi.2024}). Beyond this regime, neither laboratory validation data nor further field capacity measurements are available to assess longer-horizon lifetime predictions.
The results show how the framework developed for cell-level predictions can be adapted to mimic system-level degradation behavior. By considering the full range of stress conditions, the framework enables operators to assess the probability of an early SOH-threshold crossing, which is critical for risk-aware decision-making in maintenance planning, warranty management, and possible second-life applications.

\section{Conclusion and Outlook}
This study introduces a probabilistic framework for battery aging prediction, designed to address the inherent uncertainty in real-world degradation processes. By leveraging deep learning models trained on LFP battery datasets, the framework generates uncertainty-aware predictions of cumulative capacity loss and SOH, including prediction intervals that account for both model-related and data-driven variability. 
A key advancement of this work is its scalability to full-system data, demonstrated through comparison with multi-year field measurements from residential BESS. Hereby the framework bridges the gap between lab-scale battery modeling and real-world applications. 
This framework offers a practical tool for data-driven asset management, supporting applications such as predictive maintenance, warranty management, and possible second-life deployment. By explicitly modeling uncertainty, it enables operators to make informed, risk-aware decisions that enhance the sustainability, reliability, and economic viability of battery systems. Our framework can support aging-aware operation by identifying harmful operating regions, for example prolonged high-SOC dwelling, elevated temperatures, or unnecessarily deep and frequent cycling, and quantify the associated effect on degradation uncertainty~\cite{Collath.2023, Graner.2025}.
Looking ahead, we aim to do benchmarks with further system's operational data and furthermore incorporate forecasts of usage profiles instead of predicting SOH for defined profiles which adds a new layer of uncertainty to better reflect real-world operational dynamics.

\bibliographystyle{ieeetr}
\begin{small}  
    \bibliography{BLB}
\end{small}

\end{document}